\documentclass[runningheads]{llncs}
\usepackage[T1]{fontenc}
\usepackage{graphicx}
\usepackage{subfigure}
\usepackage{hyperref}
\usepackage[hyphenbreaks]{breakurl}

\usepackage{amsmath,amsfonts,amssymb}
\usepackage{stmaryrd}
\usepackage{graphpap}
\usepackage{graphicx} 
\usepackage{epstopdf}
\usepackage{indentfirst}
\usepackage{color}
\usepackage{listings}
\usepackage{prftree}
\usepackage{isabelle}
\usepackage{isabellesym}
\isabellestyle{it}

 \usepackage{times}
\lstdefinelanguage{isa} {alsoletter={\#, \&, $},
  mathescape=true,
  basicstyle=\small,
  escapechar=\@,
  boxpos=c,
  morekeywords= {axioms, axiomatization, Skip, Stop, type_synonym, theorem, lemma, apply, by, constdefs, definition, where,
    infixl, types, consts, primrec, have, end, show, let,  proof, qed, is,
    sorry, with, assume, fix, thus, hence, datatype, if, then, else, in, case,
    of
  },
  emph={[2] real, int, char, bool, string, self, boolean, Ass, Send, Rec, Seq, Cond, Pref,
                Join, Meet, Par, Rep, Cont, TOut, Inrp},
  emphstyle={[2]\it},
literate=
  {->}{{$\rightarrow$}}2
  {=>}{{$\Rightarrow$}}2
  {-->}{{$\Rightarrow$}}2
  {\\forall}{{$\forall$}}2
  {\\exist}{{$\exists$}}2
  {\\exists}{{$\exists$}}2
  {AND}{{$\& \&$}}3
  {ALL}{{$\forall$}}2
  {EX}{{$\exists$}}2
  {\%}{{$\lambda$}}1
  {\\\/}{{$\sqcap$}}2
  {|-}{{$\vdash$}}2
  {==}{{$\equiv$}}2
  {==>}{{$\Rightarrow$}}2
  {~}{{$\neg$}}1
  {~=}{{$\neq$}}2
  {:=}{{\tt :=}}2
  {;}{{\tt ;}}1
  {$}{{\do}}1
}

\lstnewenvironment{isaenv} {
  \lstset{extendedchars=true,columns=flexible}
  \lstset{language=isa,deletestring=[b]', basicstyle=\footnotesize}%
} {}

\newcommand{\sfsyn}[1]{\isa{#1}} 
 
\newcommand{\oomit}[1]{}
\newcommand{\seman}[1]{[\![#1 ]\!]}

\begin{document}
\title{Machine-checked executable semantics of Stateflow}
%
%
\author{Shicheng Yi\inst{1,2} \and
Shuling Wang\inst{1}\thanks{Corresponding author: wangsl@ios.ac.cn}  \and
Bohua Zhan\inst{1,2} \and
Naijun Zhan\inst{1,2}}
\authorrunning{S. Yi et al.}
%
\institute{State Key Lab. of Computer Science, Institute of Software, Chinese Academy of Sciences \and
 University of Chinese Academy of Sciences, Beijing, China}
\maketitle              
\begin{abstract}

Simulink is a widely used model-based development environment for embedded systems. Stateflow is a component of Simulink for modeling event-driven control via hierarchical state machines and flow charts. However, Stateflow lacks an official formal semantics, making it difficult to formally prove properties of its models in safety-critical applications. In this paper, we define a formal semantics for a large subset of Stateflow,   covering complex features such as hierarchical states and transitions, event broadcasts, early return, temporal operators, and so on. The semantics is formalized in Isabelle/HOL and proved to be deterministic. We implement a tactic for automatic execution of the semantics in Isabelle, as well as a translator in Python transforming Stateflow models to the syntax in Isabelle. Using these tools, we validate the semantics against a collection of examples illustrating the features we cover.
\end{abstract}

\section{Introduction}

Simulink~\cite{Simulink} is an industrial model-based design environment for embedded systems. Its component Stateflow~\cite{Stateflow} extends it with event-driven control for modelling reactive systems based on the notions of hierarchical state machines and flow charts. Stateflow inherits Simulink's capabilities including graphical modelling, efficient simulation, and code generation to implementations of  systems. However, due to the lack of formal semantics and incomplete coverage of simulation, design using Stateflow alone is insufficient for guaranteeing correctness of safety-critical systems, such as for applications in aerospace, medical services, and so on, where formal methods based rigorous semantics, analysis and verification may be required. 

There have been prior works on formal semantics and verification of Stateflow, but they consider a limited set of Stateflow features, and many of these works also lack machine-checked implementation. There are also works translating Stateflow to other formal modelling languages, but the formal  correctness of the translation is not guaranteed.
To address the above issues, this paper defines formal semantics for a large subset of Stateflow which covers the most important features, and formalizes it in the proof assistant Isabelle/HOL. Furthermore, we implement an Isabelle tactic that automatically executes the semantics, as well as a tool translating Stateflow models to Isabelle syntax. This allows us to efficiently conduct testing on Stateflow examples and compare the results with simulation within Simulink.

Stateflow is a highly complex language whose official semantics is only described informally in its Users Guide~\cite{Stateflow} (the latest versions running over a thousand pages) and through simulation within Simulink. In this paper, we define an operational semantics for Stateflow, which characterizes the effect of executing different Stateflow constructs. The definitions are compositional, preserving the hierarchical structures of the charts. We formally define data types corresponding to Stateflow charts as well as information that are modified when running the chart. Based on these, we define operational semantics for execution of composition of states, transitions, actions, and so on. These are formalized in Isabelle/HOL, with proof that the semantics is deterministic. The semantics proposed in this paper covers all rules in Appendix A, and all but one of 35 examples in Appendix B of the Stateflow Users Guide~\cite{Stateflow}.

In order to automate the execution of the semantics in Isabelle, we implement a tactic that automatically produces the result of executing the semantics for a given Stateflow model. We also implement a tool translating Stateflow graphical models in its XML format to its representation in Isabelle. This allows us to efficiently execute the semantics and compare execution results of Stateflow models against results of simulation within Simulink. We thoroughly validate the semantics we define using examples in the Stateflow Users Guide, as well as hand-crafted examples that are used to disambiguate some tricky behaviors in Stateflow. This gives us confidence in the correctness of the semantics we define.

This work also provides a semantic foundation for verification of Stateflow models against given properties, as well as for machine-checked proofs for correctness of translation from Stateflow to other formal languages and code generation to implementations of the system. On a larger scale, this work forms a part of a model-based development framework which aims to transform graphical models based on Simulink/Stateflow and AADL (Architecture Analysis \& Design Language)  to Hybrid CSP models~\cite{He94,Zhou95} for formal analysis and verification~\cite{DBLP:journals/tcs/XuWZJTZ22}, as well as code generation to SystemC implementations~\cite{DBLP:journals/tosem/YanJWWZ20}.

The remainder of this paper is organized as follows. We review related work in Section~\ref{sec:related-work}. Section~\ref{sec:introduction} gives a brief  introduction to Stateflow. Section~\ref{sec:semantics} presents the formal syntax and the operational semantics of Stateflow as implemented in Isabelle/HOL. Section~\ref{sec:engine} describes the design of automatic execution of Stateflow models according to its semantics, as well as the translation from Stateflow models to Isabelle. We describe validation on Stateflow examples in Section~\ref{sec:example}, and conclude in Section~\ref{sec:conclusion}.

\subsection{Related Work}
\label{sec:related-work}

There have been plenty of works on  semantics of Stateflow-like modeling languages. 
Statecharts, introduced by Harel~\cite{StateCharts}, is a precursor of Stateflow for modelling reactive systems, and its semantics was extensively studied~\cite{DBLP:journals/tosem/HarelN96,Statecharts_Automata,DBLP:journals/scp/Eshuis09}. One version of the semantics in terms of hierarchical automata was formalized in Isabelle/HOL \cite{Formalizing_Statecharts}. However, Stateflow is different from Statecharts in several aspects. In particular, the execution of Stateflow is  deterministic, due to assignment of priorities to parallel states and transitions, whereas Statecharts is inherently non-deterministic. Hamon presented denotational semantics~\cite{LAP1Hamon} and operational semantics~\cite{LAP2Hamon} for a subset of Stateflow. These works provide a basis for later studies. However, they miss some important features of Stateflow such as temporal operators, early return caused by event broadcasts, and so on. Furthermore, the semantics was given as mathematical definitions without formalization in proof assistants. 
Bourbouh \emph{et al.} adapted the denotational semantics to continuation-passing style, and used this to implement an interpreter and code generator for Stateflow~\cite{DBLP:conf/lpar/BourbouhGGGKT17}. Izerrouken \emph{et al.} formalized a specification of sequencing of Simulink blocks in Coq, as part of the qualification process for the \textsc{GeneAuto} code generator for Simulink~\cite{DBLP:conf/icfem/IzerroukenPT09}. It does not consider semantics for Stateflow. 

Stateflow has also been translated to other modelling languages with formal semantics and verification support. Scaife \emph{et al.} defined a safe subset of Stateflow and described the translation of the subset into Lustre for model checking~\cite{Lustre}. Cavalcanti used Circus to specify Stateflow diagrams~\cite{DBLP:journals/entcs/Cavalcanti09}. Chen \emph{et al.} translated Stateflow to CSP\# for formal analysis using the PAT model checker~\cite{Chen0LDZ12}. Jiang \emph{et al.} proposed a translation from a subset of Stateflow to UPPAAL for verification \cite{JiangSYLGGSS19}.
The above work  covers a larger subset of Stateflow and has been used on practical case studies. However, they lack a direct formalization of Stateflow semantics, and so the correctness of translation is difficult to guarantee. They also do not consider some of the more complex features in Stateflow, such as exact conditions for early return logic, graphical functions, and messages. This paper builds upon existing work of Zou \emph{et al.} on translation of Stateflow to Hybrid CSP for verification using hybrid Hoare logic. The correctness of translation is proved using UTP theory~\cite{HCSP1,DBLP:books/sp/17/ZWZ2017}, but without formalization in a theorem prover. Guo \emph{et al}. simplified this translation procedure as well as expanding the supported features~\cite{ExamplesBenchmark}.

Compared with the above works, we define an operational semantics that covers a wider range of important features in Stateflow, including exact conditions for early return logic, graphical functions, and messages. We also formalize the semantics in Isabelle/HOL, together with automatic execution of Stateflow models based on this semantics for validation and practical use. 


\section{A Brief Review of Stateflow}
\label{sec:introduction}

In this section, we first present an example of a Stateflow chart modeling a washing machine, to show how Stateflow may be used in practice. We then briefly describe the important features of Stateflow, illustrating the particularly tricky cases with examples.

\subsection{An Example of Stateflow}

\begin{figure}
  \centering
  \includegraphics[width=1\textwidth]{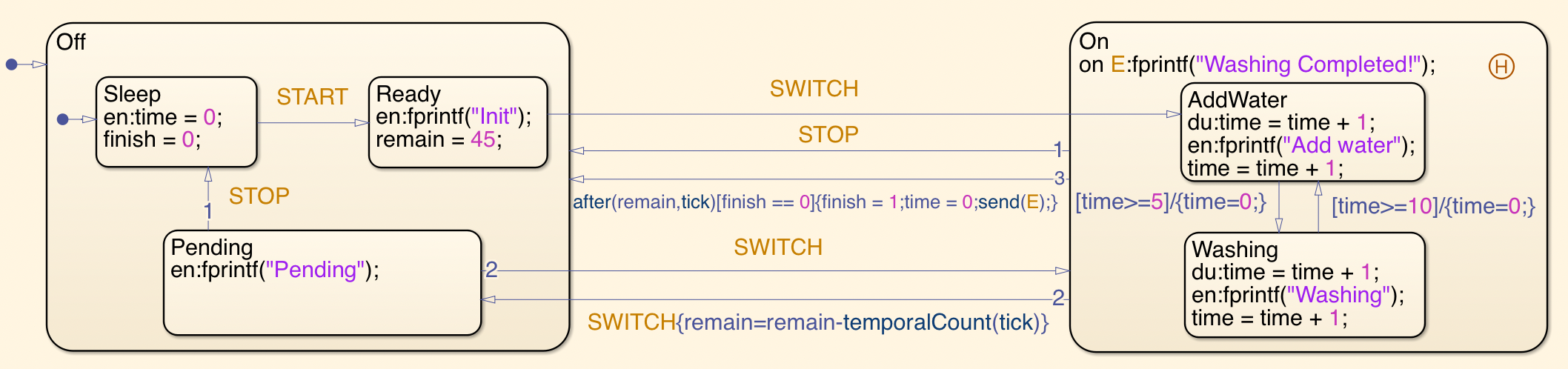} 
  \caption{A washing machine example} 
  \label{WashingMachine_Example} 
\end{figure}

Fig.~\ref{WashingMachine_Example} shows a Stateflow model for a washing machine. The washing machine has two top-level states: \sfsyn{On} and \sfsyn{Off}. The \sfsyn{Off} state is divided into three substates: \sfsyn{Sleep}, \sfsyn{Ready}, and \sfsyn{Pending}. The \sfsyn{On} state is divided into two substates: \sfsyn{AddWater} and \sfsyn{Washing}. The model has three input events: \sfsyn{START}, \sfsyn{STOP}, and \sfsyn{SWITCH}.

The washing machine starts in state \sfsyn{Off} and its substate \sfsyn{Sleep}, as indicated by the default transitions. Variables \sfsyn{finish} and \sfsyn{time} are initialized to 0 in the entry action of \sfsyn{Sleep}. The entry and during actions of a given state are defined after the symbols \sfsyn{en } and \sfsyn{du}   respectively.  Event \sfsyn{START} triggers a transition from \sfsyn{Sleep} to \sfsyn{Ready}, then event \sfsyn{SWITCH} triggers a transition from \sfsyn{Ready} to substate \sfsyn{AddWater} of \sfsyn{On}. This \emph{supertransition}, which crosses the state hierarchy, results in exit of both \sfsyn{Ready} and \sfsyn{Off}, then entry of both \sfsyn{On} and its substate \sfsyn{AddWater}.

When in state \sfsyn{On}, the washing machine alternates between staying in \sfsyn{AddWater} for 5 ticks and staying in \sfsyn{Washing} for 10 ticks. This is controlled by the variable \sfsyn{time}, which is incremented every tick in both substates, and checked/reset in the transitions between the two substates. There are three transitions from \sfsyn{On} to \sfsyn{Off}, two controlled by events and the third by execution cycles. Transition 1 is triggered by \sfsyn{STOP} to stop the machine. Transition 2 is triggered by \sfsyn{SWITCH} to pause the machine, updating \sfsyn{remain} to the remaining working time (initially 45 ticks for the washing duration), and reaches \sfsyn{Pending}, which can return to state \sfsyn{On} as soon as \sfsyn{SWITCH} is received again. A history junction is defined in state \sfsyn{On} to record the previously active substate of \sfsyn{On} before pausing the machine, which will be reentered upon receiving \sfsyn{SWITCH}. Transition 3 is triggered when the washing duration is reached, as indicated by the temporal action $\mathsf{after}(\mathit{remain},\mathsf{tick})$, where $\mathsf{tick}$ is an implicit event in Stateflow representing the execution cycles of the active states. Then, the variable $\mathit{finish}$ is set to 1 (to avoid infinite recursion of transition 3 due to event broadcast of $E$),   $\mathit{time}$ is reset to 0, and the local event $E$ is broadcast, which triggers the $\mathsf{on}~E$ action in state \sfsyn{On} to print the message $\mathtt{Washing~Completed!}$.


\subsection{Stateflow Constructs}

\subsubsection{States, junctions and transitions}

Each Stateflow chart consists of a number of \emph{states} organized in a hierarchical way. Each state may specify entry, during, and exit actions, which execute when the state is activated, remains active during a step, and becomes inactive, respectively. There are two kinds of state compositions: $\mathsf{And}$-composition for grouping parallel states and $\mathsf{Or}$-composition for grouping exclusive states. When an $\mathsf{And}$-composition becomes active, all its substates become active in a predefined order, while when an $\mathsf{Or}$-composition becomes active, only one state (specified by default transition or history junction) becomes active. In the washing machine example, both \sfsyn{On} and \sfsyn{Off} are $\mathsf{Or}$-compositions. 

\emph{Transitions} between states are specified in the form $E[c]\{a_c\}/\{a_t\}$, where $E$ is the triggering event or message, $c$ is the condition, $a_c$ and $a_t$ are the condition action and transition action respectively. When $E$ occurs and $c$ is true, the transition can be carried out, with $a_c$ executed during the transition, and $a_t$ accumulated onto a list, to be executed when a complete transition path reaching some target state is formed. \oomit{For the transitions between \sfsyn{Add\_Water}
to \sfsyn{Washing}, the transition action resetting \sfsyn{time} to 0 executes after the corresponding transition exits the source state.}Transitions originating from a state can be of two types: outer transitions and inner transitions, depending on whether the arrow leaves from the outer or inner boundary of the source state. Outer transitions are always attempted before inner transitions. Transitions crossing levels of the state hierarchy are called \emph{supertransitions} (or inter-level transitions).

\begin{figure}[h]
\centering
\subfigure{
\includegraphics[width=0.48\textwidth, height=1.6cm]{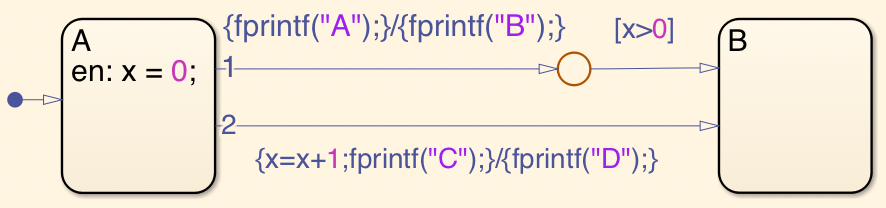}}
\subfigure{
\includegraphics[width=0.48\textwidth, height=1.6cm]{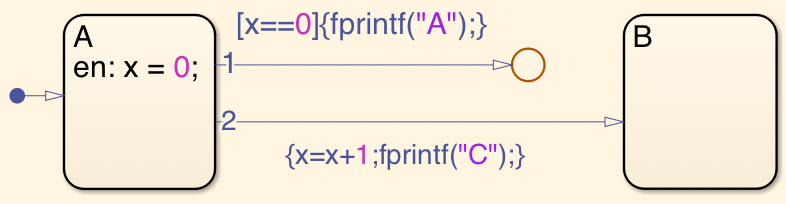}}
\caption{\textbf{(Left)}: Starting from state $A$, transition 1 is tried first, which prints $\mathtt{A}$ before failing the test $x>0$. Then transition 2 is tried, which increments $x$ and then prints $\mathtt{C}$ and $\mathtt{D}$. The visible result is printing $\mathtt{ACD}$ and reaching state $B$. \textbf{(Right)}: Junctions with no outgoing transitions present a special case. Here transition 1 is tried first, which prints $\mathtt{A}$ and reaches a terminating junction. This stops the backtracking search, so transition 2 is not tried, and state $B$ is not reached.}
\label{example:transitionpath}
\end{figure}

A \emph{junction} forms an intermediate location for transitions. They connect different transitions to form a flow chart, which can be used to represent control flows such as conditionals and loops. A transition path between two states consists of a series of transitions with junctions as intermediate points. A \emph{history junction} may be placed inside an $\mathsf{Or}$-composition to remember the previously active substate. When the $\mathsf{Or}$-composition becomes active again, the previously active substate is entered.
Fig.~\ref{example:transitionpath} shows two examples showing some of the subtleties concerning transition paths and junctions.

\subsubsection{Events and early return logic}

\begin{figure}[h]
\centering
\subfigure{
\includegraphics[width=0.5\textwidth, height=2cm]{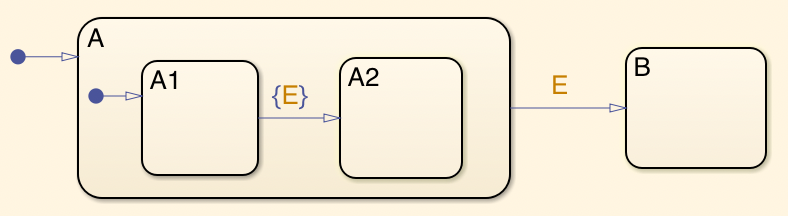}}
\subfigure{
\includegraphics[width=0.4\textwidth, height=2cm]{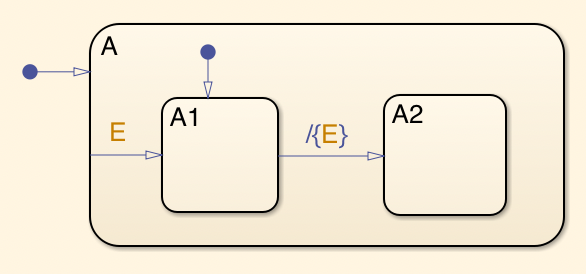}}
\caption{\textbf{(Left)}: early return for condition actions. When the transition from $A1$ to $A2$ is taken, event $E$ is broadcast, which causes the transition from $A$ to $B$. Hence, after handling $E$, states $A$ and $A1$ are no longer active, and the transition to $A2$ is abandoned. \textbf{(Right)}: early return for transition actions. The transition action of the transition from $A1$ to $A2$ is performed after exiting $A1$ and before entering $A2$. It broadcasts event $E$, which causes the re-entry of $A1$, so the transition to $A2$ is abandoned.}
\label{fig:earlyreturnlogic}
\end{figure}

Events trigger transitions or actions to occur. There are three types of events: input, output, and local events. In the washing machine example, there are three input events \sfsyn{START}, \sfsyn{STOP} and \sfsyn{SWITCH}, and one local event \sfsyn{E}. A local event is raised in actions and will cause immediate execution of its target: the entire Stateflow chart for \emph{undirected} events, or some target state in the chart for \emph{directed} events. As Stateflow charts execute in a sequential order, current activity will be interrupted to process events, and as soon as processing completes, execution continues the previous interrupted activity. Event broadcasts may cause \emph{early return}: after the event broadcast, the context for performing the remaining actions may no longer be present, so they will be discarded. Fig.~\ref{fig:earlyreturnlogic} presents the two main cases for early return logic.

\subsubsection{Other functionality}

Actions or transitions in a Stateflow chart may be guarded by events or temporal conditions. Some examples from the washing machine model are: the $\mathsf{on}~E$ guard for printing $\mathtt{Washing~Completed}$, and the guard $\mathsf{after}(\mathit{remain},\mathsf{tick})$ on transition 3 from $\mathit{On}$ to $\mathit{Off}$. Evaluating such guards necessitates keeping track of how many ticks (or seconds) the chart has stayed in any state.

\begin{figure}
\centering
\includegraphics[width=.9\textwidth]{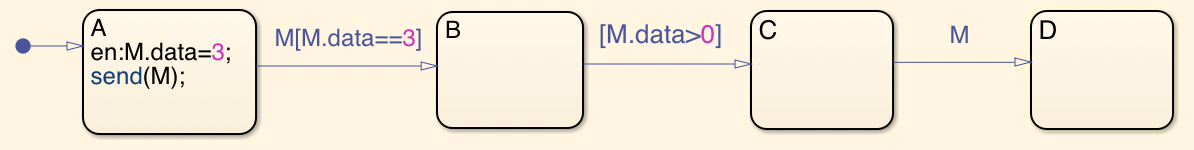} 
\caption{In state $A$, a message with name $M$ is sent with data 3. At the transition from $A$ to $B$, the event guard $M$ takes out this message, and the condition checks whether its data equals 3. The check passes so the chart transitions to $B$. At the transition from $B$ to $C$, the check is performed on the same message and passes, so the chart transitions to $C$. At the transition from $C$ to $D$, the event guard attempts to take out another message, but the queue for $M$ is empty, so the transition to $D$ will not be performed.}
\label{fig:Message} 
\end{figure}

Messages can hold data and are used to communicate between different states. After a message is sent, it is added onto a message queue according to its name. The message guard of a transition takes the top-most message from the corresponding queue, and the condition of a transition may test the content of the message (without taking new messages from the queue). Fig.~\ref{fig:Message} gives an example of using messages.

A Stateflow chart may also contain \emph{Matlab functions} and \emph{graphical functions}. A Matlab function is defined by a Matlab script, consisting of function name, a list of input variables, a list of output variables, and function body. A graphical function is similar to a Matlab function, except it is defined using a flow chart consisting of junctions. The evaluation of a graphical function largely follows that for junctions described above, except reaching a terminal junction means returning from the function.

\subsection{Execution Cycle of A State}

 When a state becomes active, it is entered first, with the consequence of executing the entry action,  then its substates (if any) are entered: all substates for And compositions and the default substate for Or compositions (if there is a history junction, the previously active substate is entered instead). As in the example, when \sfsyn{ON} is entered again due to switch, the previous substate recorded by the history function is entered. 

During the execution of a state, first the outer transitions are checked according to the priority order. If there is one transition path that is able to reach a destination state, a state transition path occurs by exiting the source state, executing the transition actions, and the target state becomes active and is entered. If a terminal junction is reached during the execution, the execution of current state stops. If there is no enabled transition path away from current state, the during and on-event actions of the state executes. Then, the inner transitions of the state is checked in priority order. If still no one is enabled, the active substates inside the current state execute recursively. 

When a state needs to be exited, first its substates are exited in the reverse order as they are entered. Then, the exit action of the state is performed.

As seen from the execution of Stateflow states explained above, entry, execution and exit of states are completely deterministic.
 
\section{Syntax and Semantics of Stateflow}
\label{sec:semantics}

We define the syntax and semantics of Stateflow. All definitions given here have been formalized in Isabelle/HOL\footnote{Implementation of the syntax and semantics, automatic tools, and examples can be found at \url{https://gitee.com/bhzhan/mars/tree/master/Semantic\_Stateflow}.}.  In this section, we  write the syntax and semantics in usual mathematical notation.

\subsection{Syntax of Stateflow Models}

We formalize the syntax of Stateflow models as follows. In the syntax,  $e$  represents expressions, $b$ Boolean variables, $E$ events, $f$ Matlab functions, and $\mathit{gf}$ graphical functions.

\vspace{-3mm}
{\small 
\[
\begin{array}{lclcl}
 \mathit{TC} & \ni & tc &:=& \mathsf{after}(n,E) ~|~ \mathsf{before}(n,E) ~|~ \mathsf{at}(n,E) ~|~ \mathsf{every}(n,E) 
 ~|~ x = \mathsf{tempCount}(E) \\
  \mathit{Cond} &\ni &	c &:=&  e_1 \textbf{ rel } e_2 ~|~ c_1 \wedge c_2 ~|~ c_1 \vee c_2 ~|~ \neg c ~|~ tc \qquad \mbox{\textbf{rel}} \in \{>, =, <\}\\
\mathit{Act} &\ni & a &:=& Skip ~|~ x::=e ~|~ ~ \mathsf{send} (E,b) ~|~ \mathsf{send}(E,b,p) ~|~ \mathsf{send}(M) ~|~ \mathsf{on}~tc::a ~|~ \mathsf{on}~E::a \\
	&&&& ~|~ \overline{x} ::= f\langle\overline{e}\rangle
  ~|~ \overline{x} ::= \mathit{gf} \langle\langle\overline{e}\rangle\rangle ~|~ \mathsf{print}(str) ~|~ a_1; a_2\\
 \mathit{Trans} &\ni & t &:= & (p_s, E, c, a_c, a_t, p_d) \qquad \qquad \ \,\mathit{TranLs}\ \ni\  tl:= \varepsilon ~|~ t\# tl\\
 \mathit{States} &\ni&\multicolumn{3}{l}{s ~|~ \mathit{None} \qquad \mathit{Juncs}\ \ni \ j ~|~ \mathit{None}\qquad \ \mathit{Paths}\ \ni \ p := \varepsilon ~|~ p.s ~|~ p.j}  \\
\mathit{SDefs} &\ni &  \multicolumn{3}{l}{sd  :=  (p, a_i, a_d, a_e, tl_i, tl_o, C) \qquad \qquad 
\mathit{JDefs} \ni  J := \{j \mapsto tl, \cdots\}}\\
\mathit{Comp} &\ni & C &:=& \mathsf{And}(L,\mathit{sf}) ~|~ \mathsf{Or}(tl,b,\mathit{sf}) \qquad \quad \mathit{SMaps}\ \ni\ \textit{sf}:=\{s \mapsto sd, \cdots\} \\

\mathit{fenv} & \ni& F& :=  & \{f_1 \mapsto (a_1, \overline{x}_1, \overline{y}_1), \cdots, f_m\mapsto (a_m, \overline{x}_m, \overline{y}_m)\} \\
\mathit{genv} & \ni & G & :=  & \{g_1 \mapsto (t_1, \overline{z}_1, \overline{r}_1), \cdots, g_k\mapsto (t_k, \overline{z}_k, \overline{r}_k)\} \\
\mathit{senv} & \ni & \Gamma   &:= & (\mathit{root}:\mathit{Comp}, F:\mathit{fenv}, G:\mathit{genv}, J:\mathit{JDefs}) 

\end{array}\]
}

A Stateflow chart $\Gamma$ (called static environment later) consists of the following parts: the state composition $\mathit{root}$ that is the root of all states in the chart; the collection of Matlab functions $F$; the collection of graphical functions $G$, and the collection of junctions $J$. Each Matlab function in $\mathit{fenv}$ has the form $f\mapsto (a,\overline{x},\overline{y})$, where action $a$ is the body of the function, and $\overline{x}, \overline{y}$ are the lists of input and output variables. Each graphical function in $\mathit{genv}$ has the form $g\mapsto (t,\overline{z},\overline{r})$, where $t$ is the initial transition, and $\overline{z},\overline{r}$ are lists of input and output variables. Each junction in $\mathit{JDefs}$ has the form $j\mapsto tl$, where $tl$ is the list of outgoing transitions from the junction.

A state composition $\mathit{Comp}$ is either an $\mathsf{And}$-composition of the form $\mathsf{And}(L,\mathit{sf})$, where $L$ is the list of names of substates in priority order, and $\mathit{sf}$ maps names to their definitions, or an $\mathsf{Or}$-composition of the form $\mathsf{Or}(tl,b,\mathit{sf})$, where $tl$ is the list of default transitions, $b$ denotes whether a history junction exists in the composition, and $\mathit{sf}$ maps names of substates to their definitions.

A state definition $sd$ is of the form $(p,a_i,a_d,a_e,tl_i,tl_o,C)$, where $p$ is the path to this state, $a_i, a_d, a_e$ are the entry, during and exit actions, $tl_i$ and $tl_o$ are the lists of inner and outer transitions in priority order, and $C$ is the internal state composition. A path $p$ is a sequence of state names, possibly ending in a junction name, indicating how to reach the state or junction starting from the root.  E.g., the path to state \sfsyn{Sleep} in Fig.~\ref{WashingMachine_Example} is \sfsyn{root.Off.Sleep}.  $\varepsilon$ represents empty path (or empty transition list in the definition of $tl$).
The path to a composition is defined as the path to its parent state. 

A transition list $tl$ is an ordered list of transitions. Each transition $t$ has the form $(p_s,E,c,a_c,a_t,p_d)$, where $p_s$ and $p_d$ are the source and destination of $t$, $E$ is the event or message guard of the transition, $c$ is the condition, and $a_c$ and $a_t$ are the condition and transition actions.

We next describe the different actions in Stateflow. The undirected event broadcast $\mathsf{send}(E,b)$ broadcasts event $E$ to the whole chart, while the directed event $\mathsf{send}(E,b,p)$ broadcasts $E$ to state composition given by path $p$. Here parameter $b$ indicates whether the sending event occurs in a transition action, to differentiate the two cases of early return in Fig.~\ref{fig:earlyreturnlogic}. Temporal action $\mathsf{on}~tc :: a$ means execution of action $a$ is guarded by the temporal condition $tc$; while $\mathsf{on}~E :: a$ means $a$ is guarded by the event $E$ or message of name $E$. Calls to Matlab functions and graphical functions are denoted by $f\langle\overline{e}\rangle$ and $\mathit{gf} \langle\langle\overline{e}\rangle\rangle$ respectively. $\mathsf{print}(str)$ prints a string $str$ (in Matlab the function is $\mathsf{fprintf}$). $a_1; a_2$ denotes sequential composition. Other control flow mechanisms, such as if-then-else and loops, are usually defined using flow charts in Stateflow.

Temporal conditions $tc$ can be event-based or absolute time based, with respective intuitive meanings. $E$ in $tc$ is either an event or specified time units.
Temporal expression $\mathsf{tempCount}(E)$ counts the number of occurrences of an event, or the number of specified time units, since the activation of the associated state. 
The syntax for conditions $c$ is as usual, with the addition of temporal conditions. 

\subsection{Configurations}

The configuration of the operational semantics includes two parts: static and dynamic environments. The static environment is simply the Stateflow chart $\Gamma$. The dynamic environment $\alpha$ has the form $(v,I)$, where $v$ contains values of variables, event and timing information, and message lists, and $I$ contains activation status and previously active substate remembered by history junctions.

\oomit{\begin{isaenv} 
	type_synonym fenv = fname => act * expl * expl
	type_synonym genv = fname => trans * expl * expl 
	type_synonym junc = path -> trans list 
	datatype senv = SEnv root fenv genv juncs 
	datatype vals = Vals var_val event_val time_val 
	datatype info = Info state_active (string option) (string option)
	type_synonym status = path -> info 
	datatype denv = DEnv vals status 
\end{isaenv}
}
\vspace{-5mm}
{\small
\[
\begin{array}{lcl}
     \mathit{vals}\ni v   &:= & (vv: \mathit{var\_val}, ev: \mathit{event\_val}, tv: \mathit{time\_val}, mv: \mathit{message\_val})  \\
     \mathit{info} \ni  i   &:= & (\mathit{is\_active}:bool, \mathit{active\_st}:\mathit{Paths}, \mathit{hj}: \mathit{Paths})\\
     \mathit{status}\ni I &:= & \{p_1 \mapsto i_1, \cdots, p_n \mapsto i_n\} \quad 
     \mathit{denv} \ni \alpha   :=  (v: \mathit{vals}, I: \mathit{status})
\end{array}\]
}
\vspace{-3mm}

The valuation $v$ has the form $(vv,ev,tv,mv)$. Here $vv$ maps variables occurring in the chart to their values; $ev$ maps path $p$ and event $e$ to the number of times that $e$ has occurred since the activation of $p$; $tv$ maps path $p$ to the simulation time that has elapsed since the activation of $p$. Finally, $\mathit{mv}$ maps message names to the corresponding message queues. The status $I$ maps each path $p$ (corresponding to a state composition) to its activation status. It consists of whether the given path is active ($\mathit{is\_active}$), the currently active substate ($\mathit{active\_st}$) and previously active substate $\mathit{hj}$ if there is a history junction. For $\mathsf{And}$-compositions, all parallel states become active or inactive together, so the latter two components are not used (always with value $\varepsilon$).



\subsection{Semantics}



The semantics of expressions $e$ is interpreted  over valuations and states, represented by $\seman{e}_{v, p}$, which returns the value of $e$ under valuation $v$ and state $p$. 
Similarly, the semantics for conditions under a given valuation $v$ and a state $p$ is defined by $\seman{c}_{v, p}$. 

The operational semantics consists of several kinds of arrows whose definitions mutually depend on each other. They range from performing a single action in the chart, to the top-level semantics for handling a sequence of events. We first explain the meaning of each of these arrows. The arrows have some common components: $\Gamma$ in the context is the Stateflow chart, $e$ on the top of the arrow indicates the current triggering event, and $\alpha_1,\alpha_2$ are the starting and ending dynamic environments, respectively. Several arrows take an additional path $p$ in the context. For actions in a state, it is the path to that state; for actions in a transition, it is the path to the source state of the transition path.
\begin{itemize}
\item $\Gamma,p\vdash (a,\alpha_1) \xrightarrow{e}_{\mathit{a}} (\alpha_2,b)$ means performing action $a$ transforms $\alpha_1$ to $\alpha_2$, and $b$ is a flag for early return: $b=\bot$ indicates early return has occurred, so the remaining actions should be abandoned, while $b=\top$ indicates early return has not occurred.

\item $\Gamma,p\vdash (t,\alpha_1) \xrightarrow{e}_{\mathit{t}} (\alpha_2,b,a_t,ts)$ means performing transition $t$ transforms $\alpha_1$ to $\alpha_2$, $b$ is the flag for early return, $a_t$ the transition action to be accumulated, and $ts$ is the target reached by the transition (either a state or a junction).

\item $\Gamma,p\vdash (tl,\alpha_1) \xrightarrow{e}_{\mathit{tl}} (\alpha_2,vt,b,\mathit{a_ts},ts,hp)$ means exploring a list of transitions $tl$ transforms $\alpha_1$ to $\alpha_2$, $vt$ indicates whether the transition has successfully reached a state, $b$ is the flag for early return, $\mathit{a_ts}$ is the accumulated transition actions, $ts$ is target state reached (if any), and $hp$ is the lowest common ancestor of states and junctions along the transition path. There are three cases for $vt$: 1 means successfully reaching a state; 0 means failing to reach a state, and $-1$ means termination due to reaching a terminal junction.

\item $\Gamma\vdash (p,\alpha_1) \xrightarrow{e}_{\mathit{exS}} (\alpha_2,b)$ means exiting from state $p$ transforms $\alpha_1$ to $\alpha_2$, with $b$ the flag for early return. $\Gamma\vdash (p,\alpha_1) \xrightarrow{e}_{\mathit{exC}} (\alpha_2,b)$ is the corresponding arrow for exiting from substates of $p$ (the composition of $p$). The arrow $\mathit{exS}$ consists of first performing $\mathit{exC}$, then calling the exit action of $p$ and exiting from $p$ itself.

\item $\Gamma,h\vdash (p,\alpha_1)\xrightarrow{e}_{\mathit{enS}} (\alpha_2,b)$ means entering state $p$ transforms $\alpha_1$ to $\alpha_2$, with $b$ the flag for early return. Here $h$ is a path (either $\varepsilon$ or starting from $p$) specifying an eventual target for entry, which is needed to define behavior of supertransitions. $\Gamma,h\vdash (p,\alpha_1)\xrightarrow{e}_{\mathit{enC}} (\alpha_2,b)$ is the corresponding arrow for entering substate of $p$ (the composition of $p$). The arrow $\mathit{enS}$ consists of first entering $p$, calling the entry action of $p$, and then performing $\mathit{enC}$.

\item $\Gamma,\mathit{is}\vdash (p,\alpha_1)\xrightarrow{e}_{\mathit{runS}} (\alpha_2,b)$ means running state $p$ transforms $\alpha_1$ to $\alpha_2$, with $b$ the flag for early return. Here $\mathit{is}$ indicates whether the current execution is in the process of handling a local event. If yes (i.e. $\mathit{is}=\top$), the simulation time on states will not be incremented. $\Gamma,\mathit{is}\vdash (p,\alpha_1)\xrightarrow{e}_{\mathit{runC}} (\alpha_2,b)$ is the corresponding arrow for running substates of $p$ (the composition of $p$).

\item $\Gamma\vdash \alpha_1\xrightarrow{[e_1,\cdots,e_n]}_{\mathit{Ch}} \alpha_2$ is the top-level arrow of the semantics, indicating that handling events $e_1$ through $e_n$ by the entire chart carries $\alpha_1$ to $\alpha_2$.

\end{itemize}

Several additional functions are used in the definition of operational semantics below. $\mathsf{lca}(p_1,\dots,p_n)$ is the least common ancestor of paths $p_1,\dots,p_n$. $\mathsf{enb}(t,\alpha,e)$ indicates whether the transition $t$ is enabled from dynamic environment $\alpha$, when the current event is $e$. $\mathsf{state}(\Gamma,p)$ returns the state definition of $p$ under $\Gamma$. $\mathsf{comp}(\Gamma,p)$ returns the composition of $p$ under $\Gamma$. The definitions of these functions are straightforward and are omitted in this paper.

We now show rules for each of the arrows in the operational semantics. For reasons of space we can only show some of the representative rules. 

\subsubsection{Semantics of expressions and conditions}

For expressions $e$, $\seman{e}_{v, p}$ returns the value of $e$ under valuation $v$ and state $p$. The state $p$ is only used for evaluating temporal expressions, as shown below for $\mathsf{tempCount}(E)$:

{\small 
\[
  \seman{\mathsf{tempCount}(E)}_{v, p} = \left\{
 \begin{array}{ll}
      v.tv(p) & \quad\mbox{if $E = \mathsf{sec}$}\\
      v.ev(p)(E) & \quad\mbox{otherwise}
\end{array}\right.
\]
}

Semantics for the conditions are standard, except for the temporal operators. These are also interpreted under a given valuation $v$ and a state $p$, defined by $\seman{c}_{v, p}$. We present some cases for conditions below.

{\small 
\begin{align*}
 \seman{\mathsf{after}(n, E)}_{v, p} = \left\{
 \begin{array}{ll}
      v.tv(p) \geq n & \quad\mbox{if $E = \mathsf{sec}$}\\
      v.ev(p)(E) \geq n & \quad\mbox{otherwise}
\end{array}\right. \\
 \seman{\mathsf{before}(n, E)}_{v, p} = \left\{
 \begin{array}{ll}
      v.tv(p) < n & \quad\mbox{if $E = \mathsf{sec}$}\\
      v.ev(p)(E) < n & \quad\mbox{otherwise}
\end{array}\right. \\
 \seman{\mathsf{at}(n, E)}_{v, p} = \left\{
 \begin{array}{ll}
      v.tv(p) = n & \quad\mbox{if $E = \mathsf{sec}$}\\
      v.ev(p)(E) = n & \quad\mbox{otherwise}
\end{array}\right. \\
 \seman{\mathsf{every}(n, E)}_{v, p} = \left\{
 \begin{array}{ll}
      v.tv(p)\ \mathrm{mod}\ n = 0 & \quad\mbox{if $E = \mathsf{sec}$}\\
      v.ev(p)(E)\ \mathrm{mod}\ n = 0 & \quad\mbox{otherwise}
\end{array}\right.
\end{align*}
}

\subsubsection{Semantics of actions}
  
\begin{figure}[h!]
	{\small 
		\begin{eqnarray*} 
			&  \prftree[r]{SendF}
			{\Gamma, \top \vdash (\Gamma[1],  \alpha_1)  \xrightarrow{e'}_{\mathit{exeC}} ((v_2, I_2), b) \quad I_2(p) = (b_1, p_a, p_h)}
			{\Gamma, p \vdash (\mathsf{send}(e', 0), \alpha_1) \xrightarrow{e}_a ((v_2, I_2), b_1) } & \\
			&  \prftree[r]{SendT}
			{\Gamma, \top \vdash (\Gamma[1], \alpha_1)  \xrightarrow{e'}_{\mathit{exeC}} ((v_2, I_2), b) \quad I_2 \ (\mathsf{parent}(p)) = (b_1, p_a, p_h) }
			{\Gamma, p \vdash (\mathsf{send}(e', 1), \alpha_1) \xrightarrow{e}_a ((v_2, I_2), (b_1 \wedge p_a = \varepsilon)) }&\\
			&	 \prftree[r]{SendM}
			{
			}
			{\Gamma, p \vdash (\mathsf{send}(M), (v, I)) \xrightarrow{e}_a ((v(v.mv[M \mapsto v.mv(M)@\seman{M.data}_{v,p}]), I), \top)} & \\
			& \prftree[r]{OnT}
			{\begin{array}{cc}
				\seman{tc}_{v, p}	  = \top \\
					 \Gamma, p \vdash (a, \alpha) \xrightarrow{e}_a \gamma 
				\end{array}  }
			{\Gamma, p \vdash (\mathsf{on}\ tc :: a, \alpha) \xrightarrow{e}_a \gamma   }
			\quad
			\prftree[r]{OnF}
			{\begin{array}{cc}
					\seman{tc}_{v, p} = \bot 
			\end{array}  }
			{\Gamma, p \vdash (\mathsf{on }\ tc :: a, \alpha) \xrightarrow{e}_a (\alpha, \top) }&\\
			&\prftree[r]{OnE}
			{\begin{array}{cc}
					e = E \quad \Gamma, p \vdash (a, \alpha) \xrightarrow{e}_a \gamma 
			\end{array}  }
			{\Gamma, p \vdash (\mathsf{on }\ E :: a, \alpha) \xrightarrow{e}_a \gamma  }&\\
			&\prftree[r]{SeqT}
			{\begin{array}{cc}
					\Gamma, p \vdash (c_1, \alpha_1) \xrightarrow{e}_a (\alpha_2, \top) \\
					\Gamma, p \vdash (c_2, \alpha_2) \xrightarrow{e}_a \gamma 
			\end{array}  }
			{\Gamma, p \vdash (c_1; c_2, \alpha_1) \xrightarrow{e}_a \gamma   }
			\quad
			 \prftree[r]{SeqF}
			 {\begin{array}{cc}
			 		\Gamma, p \vdash (c_1, \alpha_1) \xrightarrow{e}_a (\alpha_2, \bot)
			 \end{array}  }
			 {\Gamma, p \vdash (c_1; c_2, \alpha_1) \xrightarrow{e}_a (\alpha_2, \bot)  }	& \\
			 & \prftree[r]{GraF}
			 {\begin{array}{cc}
			 		\Gamma.G(\mathit{gf}) = (t, \overline{y}, \overline{z})  \quad \Gamma, \varepsilon \vdash ([t], (v_1[\overline{y} \mapsto \seman{\overline{w}}_{v_1, p}], I_1)) \xrightarrow{e}_{\mathit{tl}} ((v_2, I_2), -1, \top, \_, \_, \_)  
			 \end{array}  }
			 {\Gamma, p \vdash (\overline{x}::=\mathit{gf}\langle\langle \overline{w} \rangle \rangle, (v_1, I_1)) \xrightarrow{e}_a ((v_2[\overline{x} \mapsto v_2(\overline{z})], I_2), \top) }	& \\
			 & \prftree[r]{MatF}
			 {\begin{array}{cc}
			 		\Gamma.F(mf) = (c, \overline{y}, \overline{z})  \quad \Gamma, \varepsilon \vdash (c, (v_1[\overline{y} \mapsto v_1(\overline{w})], I_1)) \xrightarrow{e}_{\mathit{a}} ((v_2, I_2), \top)  
			 \end{array}  }
			 {\Gamma, p \vdash (\overline{x}::=\mathit{mf}\langle \overline{w} \rangle, (v_1, I_1)) \xrightarrow{e}_a ((v_2[\overline{x} \mapsto v_2(\overline{z})], I_2), \top) }	&
		\end{eqnarray*}
	}
\caption{Semantics rules for actions}
\end{figure}

For the semantics of actions, rules (SendF) and (SendT) broadcast $e'$ to the root of the chart,  for the cases on whether or not event broadcast occurs in transition action. In consequence, the top composition $\mathit{root}$ executes under the context of handling local event $e'$. The resulting status $I_2$ is used for deciding early return logic: for (SendF), if  $p$ is still active, i.e. $b_1=\top$, then the remaining actions are continued; for (SendT), if the parent state of $p$, denoted by $\mathsf{parent}(p)$, is active, and all substates inside $\mathsf{parent}(p)$ are inactive, indicated by $p_a = \varepsilon$, then the remaining actions are continued. Rule (SendM) defines the semantics of sending a message, which adds the message value to its
queue. 

Rules (OnT) and (OnF) check the truth of temporal condition $\mathit{tc}$, and proceed with $a$ if it is true, otherwise not. Rule (OnE) defines when $E$ is received, $a$  executes. This is the only type of actions triggered by events. 

For sequential composition, it will check the value of the flag for early return logic, if it is true,  $c_2$ continues to execute (rule SeqT), otherwise, an early return occurs and $c_2$ will be discarded  (rule SeqF).

Rule (GraF) defines the semantics for executing a graphical function. Suppose $\Gamma.G(\mathit{gf})$ has the form $(t,\overline{y},\overline{z})$. The call to $\mathit{gf}$ is equivalent to first assigning input variables $\overline{y}$ to their respective values, then executing the transition list $[t]$ (i.e. the flow chart of $\mathit{gf}$), and finally assigning values of output variables to $\overline{x}$. We use ``$\_$'' to denote values in the tuple that are unused, or values that are unchanged in an assignment.

The semantics for executing a Matlab function (rule MatF) is defined similarly, where the main process is to execute the function body $c$ as a action.

\begin{example}
The local event broadcast on transition 3 from \sfsyn{On} to \sfsyn{Off} in Fig.~\ref{WashingMachine_Example} is represented in our syntax by $\mathsf{send}(E,0)$, where $0$ stands for condition action. The (rule SendF) is applied, causing the execution of the entire chart using the arrow $\mathit{runC}$. The arrow outputs $\mathtt{Washing~Completed!}$ but does not result in change of activation status of states, so state $\sfsyn{On}$ is still active ($b_1=\top$), and there is no early return.
 \label{example:send}
\end{example}


\subsubsection{Semantics of transitions and transition lists}

\begin{figure}[h!]
	{\small 
		\begin{eqnarray*} 
		    &  \prftree[r]{Updv}
		    {\begin{array}{cc}
		        E = \mathsf{Message} \ m \quad v.mv(m) \neq \ [\ ]
		     \end{array}
		    }
		    {v \rightarrow_E v(v.vv[m \mapsto (\mathsf{head}(v.mv(m)))], v.mv[m \mapsto (\mathsf{tail}(v.mv(m)))])
		    } &\\
			&  \prftree[r]{TrT}
			{\begin{array}{cc}
			     t = (p_s,E,c,a_c,a_t,p_d) \\
				 \mathsf{enb}(t, \alpha_1, e) \quad v_1 \rightarrow_E v_2\\
				 \Gamma, p \vdash (a_c, (v_2, I_1)) \rightarrow_a (\alpha_2, \top)
				\end{array}
			}
			{\Gamma, p \vdash (t, (v_1, I_1)) \xrightarrow{e}_{\mathit{t}} (\alpha_2, \top, a_t, d)} 
		\quad  \prftree[r]{TrF}
			{\begin{array}{cc}
			        t = (p_s,E,c,a_c,a_t,p_d) \\
					\mathsf{enb}(t, \alpha_1, e)  \quad v_1 \rightarrow_E v_2 \\
					\Gamma, p \vdash (a_c, (v_2, I_1)) \rightarrow_a (\alpha_2, \bot)
				\end{array}
			}
			{\Gamma, p \vdash (t, (v_1, I_1)) \xrightarrow{e}_{\mathit{t}} (\alpha_2, \bot, \varepsilon, \mathit{None})}&\\	
			&	 \prftree[r]{Emp}
			{ 
			}
			{\Gamma, p \vdash (\varepsilon, \alpha) \xrightarrow{e}_{\mathit{tl}} (\alpha, -1, \top, \varepsilon, \mathit{None}, \varepsilon)}\\ 
			&\prftree[r]{ToS}
			{ \Gamma, p \vdash (t, \alpha_1) \xrightarrow{e}_{t} 
				(\alpha_2, \top, a, d) \quad d \in \mathit{States}
			}
			{\Gamma, p \vdash (t\#tl, \alpha_1) \xrightarrow{e}_{\mathit{tl}}   (\alpha_2, 1, \top, a, d, \mathsf{lca}(s, d))} &\\
			&\prftree[r]{ToHJ}
			{ \begin{array}{c}
					\Gamma, p \vdash (t, \alpha_1) \xrightarrow{e}_{t} 
					((v_2, I_2), \top, a, d) \quad d \mbox{ is a history junction } \quad I_2(d) = (b',p_a,p_h)   
				\end{array}
			}
			{\Gamma, p \vdash (t\#tl, \alpha_1) \xrightarrow{e}_{\mathit{tl}}  ((v_2, I_2), 1, \top, a, p_h, \mathsf{lca}(s, d))  } &\\
			&\prftree[r]{ToJ1}
			{ \begin{array}{c}
					\Gamma, p \vdash (t, \alpha_1) \xrightarrow{e}_{t} 
					(\alpha_2, \top, a_1, d) \quad d \in \mathit{Juncs} \\
					\Gamma, p \vdash (\Gamma.J(d), \alpha_2) \xrightarrow{e}_{\mathit{tl}}   (\alpha_3, 1, \top, a_2, d_2, p_2) 
				\end{array}
			}
			{\Gamma, p \vdash (t\#tl, \alpha_1) \xrightarrow{e}_{\mathit{tl}}   (\alpha_3, 1, \top, (a_1;a_2), d, \mathsf{lca}(s, d, p_2))} &\\
			&\prftree[r]{ToJ2}
			{ \begin{array}{c}
				tl \neq \varepsilon \quad 	\Gamma, p \vdash (t,\alpha_1) \xrightarrow{e}_{t} 
					(\alpha_2, \top, a_1, d) \quad d \in \mathit{Juncs} \\   
					\Gamma, p \vdash (\Gamma.J(d), \alpha_2) \xrightarrow{e}_{\mathit{tl}}   (\alpha_3, 0, \top, \varepsilon, \mathit{None}, \varepsilon)  
					\quad \Gamma, p \vdash (tl, \alpha_3) \xrightarrow{e}_{\mathit{tl}} \gamma
				\end{array}
			}
			{\Gamma, p \vdash (t\#tl, \alpha_1) \xrightarrow{e}_{\mathit{tl}}  \gamma } & \\	
			&\prftree[r]{ToJ3}
			{ \begin{array}{c}
					\Gamma, p \vdash (t, \alpha_1) \xrightarrow{e}_{t} 
					(\alpha_2, \top, a_1, d) \quad d \in \mathit{Juncs} \\
					\Gamma, p \vdash (\Gamma.J(d), \alpha_2) \xrightarrow{e}_{\mathit{tl}}   (\alpha_3, 0, \top, \varepsilon, \mathit{None}, \varepsilon)   
				\end{array}
			}
			{\Gamma, p \vdash ([t], \alpha_1) \xrightarrow{e}_{\mathit{tl}}  (\alpha_3, 0, \top, \varepsilon, \mathit{None}, \varepsilon)    } &\\	
			&\prftree[r]{ToJ4}
			{ \begin{array}{c}
					\Gamma, p \vdash (t, \alpha_1) \xrightarrow{e}_{t} 
					(\alpha_2, \top, a_1, d) \quad d \in \mathit{Juncs} \\
					\Gamma, p \vdash (\Gamma.J(d), \alpha_2) \xrightarrow{e}_{\mathit{tl}}   (\alpha_3, -1, \top, \varepsilon, \mathit{None}, \varepsilon)   
				\end{array}
			}
			{\Gamma, p \vdash (t\#tl, \alpha_1) \xrightarrow{e}_{\mathit{tl}}  (\alpha_3, -1, \top, \varepsilon, \mathit{None}, \varepsilon)  } & \\	
			& \prftree[r]{Ind}
			{\begin{array}{cc}
			\neg \mathsf{enb}(t, \alpha_1, e) \quad v_1 \rightarrow_E v_2 \\ 
			tl \neq \varepsilon \quad
				\Gamma, p \vdash (tl, (v_2, I_1)) \xrightarrow{e}_{tl}   \gamma 	
				\end{array}
			}
			{\Gamma, p \vdash (t\#tl, (v_1, I_1)) \xrightarrow{e}_{\mathit{tl}}  \gamma}  \quad 
			\prftree[r]{Fail}
			{ \neg \mathsf{enb}(t, \alpha_1, e)  \quad v_1 \rightarrow_E v_2
			}
			{\Gamma, p \vdash ([t], (v_1, I_1)) \xrightarrow{e}_{\mathit{tl}}   ((v_2, I_1), 0, \top, \varepsilon, \mathit{None}, \varepsilon)}	&\\					 					 	
		\end{eqnarray*}
	}
\caption{Semantics rules for transitions}
\end{figure}

For the rules of transitions, a transition $(p_s, E, c, a_c, a_t, p_d)$ is enabled under $\alpha_1=(v_1,I_1)$ and event $e$, denoted by $\mathsf{enb}(t,\alpha_1,e)$, if $\seman{c}_{v_1, p_s}$ holds, and $ E=e \vee E= \varepsilon \vee v_1.mv(E) \neq [\ ]$ holds. The arrow $v_1 \rightarrow_E v_2$ is defined as (rule Updv) to pop a message from the message queue ($v_1.mv$) and record the message value ($v_1.vv$) if $E$ is a message. Then
when transition $t$ is enabled, $a_c$ will be executed, then if the execution returns with $b=\top$ (early return does not occur), the transition action $a_t$ and target $d$ are recorded (rule TrT), otherwise not (rule TrF). 

We next list rules for execution of a transition list. Suppose the transition list is in the form $t\#tl$. If $t$ is enabled and reaches state $d$, then the execution of the transition list completes, with $vt=1$, $hp$ the lowest common ancestor of source $p$ and target $d$, i.e. $\mathsf{lca}(p, d)$ (rule ToS). If $t$ is enabled but reaches a junction $d$, then repeat the process on the outgoing transition list of $d$, i.e. $\Gamma.J(d)$. If the outgoing transitions of $d$ finally reaches a state (returned $vt$ is 1), a complete transition path is found (rule ToJ1). If the outgoing transitions of $d$ fail to reach a state (returned $vt$ is 0), then backtrack to the previous transition list $tl$ to execute (rule ToJ2).
But if $tl$ is empty, the whole execution terminates and fails to reach a state (rule ToJ3). If the outgoing transitions of $d$ reaches a terminal junction (returned $vt$ is $-1$), the whole execution is recorded as reaching a terminal junction (rule ToJ4). If $t$ is not enabled, we update $v$ and repeat the process on the rest of the transition list $tl$ (rule Ind). But if $tl$ is empty,  the whole execution fails directly (rule Fail).


\oomit{
\begin{example}
Continue Example~\ref{example:send} for illustration, after the event broadcast finishes, the execution returns back to transition 3, whose execution will result in $(v_3, I_3, 1, \varepsilon, \sfsyn{OFF})$ for some $v_3$ and $I_3$ such that the time of each state reset to 0, and $I_3$ turns to 

\vspace{-5mm}
{\small
\[
\begin{array}{ll}
   \{\mathit{root} \mapsto (1, OFF, None), \mathit{root.OFF} \mapsto (1, \mathit{off}, \_), \mathit{root.ON} \mapsto (0, None, None), \cdots\}  
\end{array}
\]
}
\end{example}
\vspace{-5mm}
}
\begin{example}
Fig.~\ref{example:transitionpath} (left) shows an example of  backtracking. Starting from state $A$, transition 1 is tried first and reaches a junction. Since the transition following the junction cannot execute, it returns $vt=0$. This causes backtracking, and transition 2 is tried, which reaches state $B$ and returns $vt=1$ (rule ToS), so executing the whole transition list reaches $B$ and returns $vt=1$ by (rule ToJ2).

Fig.~\ref{example:transitionpath} (right) shows an example of  stopping due to reaching a terminal junction. Starting from state $A$, transition 1 is tried first and reaches the junction, but there is no outgoing transitions from the junction, so it returns $vt=-1$. This causes execution of the whole transition list to return $vt=-1$ according to (rule ToJ4).
\end{example}

\subsubsection{Semantics of state and composition exit and entry}

\begin{figure}[h!]
	{\small 
		\begin{eqnarray*} 
			&  \prftree[r]{exS1}
			{\begin{array}{cc}
			\Gamma \vdash (p, \alpha_1) \xrightarrow{e}_{\mathit{exC}} (\alpha_2, \top)	\\
			\mathsf{state}(\Gamma, p) = (p, a_i, a_d, a_e, tl_i, tl_o, C) \\ 
			 \Gamma, p \vdash (a_e, \alpha_2) \xrightarrow{e}_{\mathit{a}} ((v_3, I_3), \top) 
				\end{array}
			}
			{\Gamma \vdash (p, \alpha_1) \xrightarrow{e}_{\mathit{exS}} ((v_3, I_3[p \mapsto (\bot, \_, \_)]), \top)} 
			  \quad \prftree[r]{exS2}
			 {\begin{array}{cc}
			 		\Gamma \vdash (p, \alpha_1) \xrightarrow{e}_{\mathit{eC}} (\alpha_2, \bot) \\
			 	\end{array}
			 }
			 {\Gamma \vdash (p, \alpha_1) \xrightarrow{e}_{\mathit{eS}} (\alpha_2, \bot)} 
			 &\\
			 &  \prftree[r]{exS3}
			{\begin{array}{cc}
			\Gamma \vdash (p, \alpha_1) \xrightarrow{e}_{\mathit{exC}} (\alpha_2, \top)	\\
			\mathsf{state}(\Gamma, p) = (p, a_i, a_d, a_e, tl_i, tl_o, C) \quad 
			 \Gamma, p \vdash (a_e, \alpha_2) \xrightarrow{e}_{\mathit{a}} (\alpha_3, \bot) 
				\end{array}
			}
			{\Gamma \vdash (p, \alpha_1) \xrightarrow{e}_{\mathit{exS}} (\alpha_3, \bot)} &\\
			 &\prftree[r]{exO}			 {\begin{array}{cc}
			 		\mathsf{comp}(\Gamma,p) = \mathsf{Or}(\mathit{tl}, b, \textit{sf}) \quad I_1(p) = (b',p_a,p_h) \quad \Gamma \vdash (p_a, (v_1, I_1)) \xrightarrow{e}_{\mathit{exS}} ((v_2, I_2), \top) \\
			 	  I_3 = I_2[p \mapsto (\_,\varepsilon, \_)]
			 	  \quad 	b \rightarrow I_4 = I_3[p \mapsto (\_, \_, p_a)] \quad \neg b \rightarrow I_4 = I_3 
			 	\end{array}
			 }
			 {\Gamma\vdash (p, (v_1, I_1)) \xrightarrow{e}_{\mathit{exC}} ((v_2, I_4), \top)}
			 &\\
			 &\prftree[r]{exSL}
			 {\begin{array}{cc}
			 		\Gamma \vdash (s, \alpha_1) \xrightarrow{e}_{\mathit{exS}} (\alpha_2, \top) \\
			 		\Gamma \vdash (sl, \alpha_2) \xrightarrow{e}_{\mathit{exSL}} \gamma 
			 	\end{array}
			 }
			 {
			    \Gamma \vdash (sl@[s], \alpha_1) \xrightarrow{e}_{\mathit{exSL}} \gamma} 
			 \quad \prftree[r]{exA}
			 {\begin{array}{cc}
			        \mathsf{comp}(\Gamma,p) = \mathsf{And}(\mathit{sl}, \textit{f}) \\
			 		\Gamma \vdash (sl, \alpha_1) \xrightarrow{e}_{\mathit{exSL}} ((v_2, I_2), \top) \\
			 		I_3 = I_2[p \mapsto (\bot,\varepsilon,  \varepsilon) ] 
			 	\end{array}
			 }
			 {\Gamma \vdash (p, \alpha_1) \xrightarrow{e}_{\mathit{exC}} (v_2, I_3, \top)} 
          &\\						 	
		\end{eqnarray*}
	}
\caption{Semantics for exiting from states}
\end{figure}

After a transition completes successfully, the source state exits and the target state is entered. Whenever a state is entered or exited, the activation status of the state, its substates, some of its superstates, as well as their sibling states will be changed (the latter two in the case of supertransitions).

Given state $s = (p, a_i, a_d, a_e, ti, to, C)$, (rule exS1) defines how to exit from state $p$: first exit the composition $C$ of $p$, then execute the exit action $a_e$ of $p$, and finally update the status of $p$ to be inactive. If early return occurs in the exit of composition $C$, the whole execution terminates immediately (rule exS2). If early return occurs in the execution of $a_e$, the remainder of the execution is abandoned as well (rule exS3). we will omit some rules related to early return in the following.

(Rule exO) defines how to exit from an $\mathsf{Or}$-composition: first exit from the active substate of $C$, i.e. $p_a$, then update the active substate of $C$ to be empty, and if the flag $b$ in the composition is true, indicating presence of history junction, records the previously active substate $p_a$. (Rule exA) defines the exiting of an $\mathsf{And}$-composition, which exits the parallel states in the reverse order with respect to their priority (defined by rule exSL),  then the context is updated.

\begin{figure}[h!]
	{\small 
		\begin{eqnarray*} 
			&  \prftree[r]{enS}
			{
				\begin{array}{cc}
					v_2 = v_1[\_, p \mapsto \lambda ev.\,0, p \mapsto 0, \_] \quad 
					I_2 = I_1[p \mapsto (\top, \_, \_) , \mathsf{parent}(p) \mapsto (\_, p,\_)] \\
					\mathsf{state}(\Gamma, p) = (p, a_i, a_d, a_e, tl_i, tl_o, C) \\
					\Gamma, p \vdash (a_i, (v_2, I_2)) \xrightarrow{e}_{a}(\alpha_3, \top) \quad  
					\Gamma, \mathsf{tail}(h) \vdash (p, \alpha_3) \xrightarrow{e}_{\mathit{enC}} (\alpha_4, b)
				\end{array}
			}
			{\Gamma, h \vdash (p, (v_1, I_1)) \xrightarrow{e}_{\mathit{enS}} (\alpha_4, b)}  
			&\\
			 &  \prftree[r]{enO1}
			 {
			 	\begin{array}{cc}
			 		\mathsf{comp}(\Gamma, p) = \mathsf{Or}(\mathit{tl}, b, \textit{sf}) \quad h \neq \varepsilon  \quad
			 		\Gamma, h \vdash (p.\mathsf{head}(h), \alpha_1) \xrightarrow{e}_{\mathit{enS}}(\alpha_2, b_1)
			 	\end{array}
			 }
			 {\Gamma, h \vdash (p, \alpha_1) \xrightarrow{e}_{\mathit{enC}} (\alpha_2, b_1)} 
			 &\\
			 	&  \prftree[r]{enO2}
			 	{
			 		\begin{array}{cc}
			 		\mathsf{comp}(\Gamma, p) = \mathsf{Or}(\mathit{tl}, b, \mathit{sf}) \quad
			 		h = \varepsilon \quad b = \top \\ 
			 			I_1(p) = (b', p_a, p_h) \quad p_h \neq  \varepsilon \quad
			 			\Gamma, h \vdash (p_h, (v_1, I_1)) \xrightarrow{e}_{\mathit{enS}}(\alpha_2, b_1)
			 		\end{array}
			 	}
			 	{\Gamma, h \vdash (p, (v_1, I_1)) \xrightarrow{e}_{\mathit{enC}} (\alpha_2, b_1)} 
			 &\\
			 &  \prftree[r]{enO3}
			 {
			 	\begin{array}{cc}
			 		\mathsf{comp}(\Gamma, p) = \mathsf{Or}(\mathit{tl}, b, \mathit{sf})\quad h = \varepsilon \quad (b = \top \wedge
			 		I_1(p) = (b', p_a, p_h) \wedge p_h = \varepsilon) \vee b = \bot \\
			 		\Gamma, p	\vdash (tl, (v_1, I_1)) \xrightarrow{e}_{\mathit{tl}} (\alpha_2, \_, \top, a_t, ts, \_) \quad 
			 		\Gamma, p\vdash (a_t, \alpha_2) \xrightarrow{e}_{\mathit{a}} (\alpha_3, \top) \\
			 		\Gamma, ts\backslash p \vdash (ts, \alpha_3) \xrightarrow{e}_{\mathit{enS}}(\alpha_4, b_1)
			 	\end{array}
			 }
			 {\Gamma, h \vdash (p, (v_1, I_1)) \xrightarrow{e}_{\mathit{enC}} (\alpha_4, b_1)} 
			 &\\
			 &  \prftree[r]{enSL}
			 {
			 	\begin{array}{cc}
			 		h' = (\textbf{if } s = \mathsf{head}(h)\textbf{ then } h \textbf{ else } \varepsilon) \\ \Gamma, h' \vdash (s, \alpha_1) \xrightarrow{e}_{\mathit{enS}} (\alpha_2, \top) \quad
			 		\Gamma, h \vdash (sl, \alpha_2) \xrightarrow{e}_{\mathit{enSL}}(\alpha_3, b)
			 	\end{array}
			 }
			 {\Gamma, h, f \vdash (s\#sl, \alpha_1) \xrightarrow{e}_{\mathit{enSL}}(\alpha_3, b)} 
			 &\\	
			 &  \prftree[r]{enA}
			 {
			 	\begin{array}{cc}
			 	    \mathsf{comp}(\Gamma, p) = \mathsf{And}(\mathit{sl}, \textit{f}) \quad
			 		\Gamma, h\vdash (sl, \alpha_1) \xrightarrow{e}_{\mathit{enSL}}(\alpha_2, b)
			 	\end{array}
			 }
			 {\Gamma, h \vdash (p, \alpha_1) \xrightarrow{e}_{\mathit{enC}} (\alpha_2, b)} 
			 &\\					 	
		\end{eqnarray*}
	}
	\caption{Semantics for entering into states}
\end{figure}

The semantics of entering states and compositions is more complicated with some extra tasks. The event and time valuations for a state need to be reset at activation. For entry into a composition, if it is part of performing a supertransition where which substate should be entered is known, the given substate is entered. Otherwise, the substate to be entered is determined by the default transitions or the history junction if present. Recall the parameter $h$ in the context indicates the eventual target of entry when performing a supertransition.

When state $p$ is entered (rule enS): (1) the event and time valuations of $p$ are reset to 0; (2) the state $p$ becomes active, and it becomes the active substate of the parent of $p$; (3) the entry action $a_i$ of $p$ executes; (4) the composition $C$ is entered, where the path from $C$ to the target becomes the tail of the input path $h$, i.e. $\mathsf{tail}(h)$.

For entry into an $\mathsf{Or}$-composition, there are three different cases depending on whether $h$ is empty: if $h$ is not empty, then the first substate recorded in path $h$ is entered (rule enO1);  if $h$ is empty, and if the composition has stored a previously active substate $p_h$, then $p_h$ is entered (rule enO2); otherwise, the default transition list $tl$ will execute and then the target reached by $tl$, that is $ts$, is chosen to be entered (rule enO3). For  $\mathsf{And}$-composition, the parallel states are entered in the priority order, with two different cases depending on whether the target to be entered is inside the states or not (rules enSL, enA).

\begin{example}
We use the washing machine example to demonstrate the exit and entry of states and compositions. Suppose state \sfsyn{On} and its substate \sfsyn{Washing} are active, and event \sfsyn{SWITCH} occurs. Then transition 2 from \sfsyn{On} to \sfsyn{Pending} executes. According to the rules, the following entry and exit actions are taken in sequence: (1) state \sfsyn{Washing} exits; (2) state \sfsyn{On} exits; (3) state \sfsyn{Off} is entered using (rule enO1) with $h$ being \sfsyn{Off.Pending}; (4) state \sfsyn{Pending} is entered using (rule enO1) with $h$ being \sfsyn{Pending}.

If another \sfsyn{SWITCH} occurs, transition 1 starting from \sfsyn{Pending} is executed, reaching target state \sfsyn{On}. Then state \sfsyn{On} is entered, followed by entering state \sfsyn{Washing} using (rule enO2) since \sfsyn{Washing} is recorded as the previously active substate.

If event \sfsyn{STOP} occurs while in state \sfsyn{On}, then
transition 1 from \sfsyn{On} to \sfsyn{Off} is executed. This causes entry of state \sfsyn{Off} and then state \sfsyn{Sleep} by the default transition, using (rule enO3).
\end{example}

\subsubsection{Semantics of state execution}

\begin{figure}[h!]
	{\small 
		\begin{eqnarray*} 
			&  \prftree[r]{runS}
			{
				\begin{array}{cc}
				v_2.ev = v_1.ev[(p, e)  \mapsto v_1.ev(p,e)+1] \quad
				v_3 = \textbf{if}\ is \ \textbf{then}\ v_2\ \textbf{else}\ v_2.tv[p \mapsto v_2.tv(p)+1] \\
			\mathsf{state}(\Gamma, p) = (p, a_i, a_d, a_e, tl_i, tl_o, C) \quad 	\Gamma, p \vdash (tl_o, (v_3, I_1)) \xrightarrow{e}_{tl} (\alpha_2, 1, \top, a_t, ts, hp)	 \\
				exS = \textbf{if}\ (p = ts = hp) \ \textbf{then}\ \mathsf{parent}(p) \ \textbf{else}\ \mathsf{lca}(p, hp)\\
				\Gamma \vdash (exS, \alpha_2) \xrightarrow{e}_{\mathit{exC}} (\alpha_3, \top) \quad 
				\Gamma, p \vdash (a_t, \alpha_3) \xrightarrow{e}_{\mathit{a}} (\alpha_4, \top) \\
				enS = \textbf{if}\ (p = ts = hp) \ \textbf{then}\  ts.\mathit{parent} \ \textbf{else}\ \mathsf{parent}(ts) \\
				h = \textbf{if}\ (p = ts = hp) \ \textbf{then}\  [\mathsf{last}(ts)] \ \textbf{else}\ ts\backslash hp\quad
				\Gamma, h\vdash (enS, \alpha_4) \xrightarrow{e}_{\mathit{enC}} (\alpha_5, b)
				\end{array}
			}
			{\Gamma, is  \vdash (p, (v_1, I_1)) \xrightarrow{e}_{\mathit{runS}} (\alpha_5, b)}
			&\\	
			&  \prftree[r]{runS2}
			{
				\begin{array}{cc}
					v_2.ev = v_1.ev[(p, e)  \mapsto v_1.ev(p,e)+1] \quad  
				v_3 = \textbf{if}\ is \ \textbf{then}\ v_2\ \textbf{else}\ v_2.tv[p \mapsto v_2.tv(p)+1] \\
				\mathsf{state}(\Gamma, p) = (p, a_i, a_d, a_e, tl_i, tl_o, C) \\
					\Gamma, p \vdash (tl_o, (v_3, I_1)) \xrightarrow{e}_{\mathit{tl}} (\alpha_2, b, \top, a_t, ts, hp)\quad b=0\vee b=-1	 \\
					\Gamma, p \vdash (a_d, \alpha_2) \xrightarrow{e}_{\mathit{a}} (\alpha_3, \top) \quad 
					\Gamma, p \vdash (tl_i, \alpha_3) \xrightarrow{e}_{\mathit{tl}} (\alpha_4, 1, \top, a_t', ts', hp')\\
					exS = \mathsf{lca}(p, hp) \quad
					\Gamma \vdash (exS, \alpha_4) \xrightarrow{e}_{\mathit{exC}} (\alpha_5, \top) \\
					\Gamma, p \vdash (a_t', \alpha_5) \xrightarrow{e}_{\mathit{a}} (\alpha_6, \top) \quad
					enS = \textbf{if}\ (p = ts = hp) \ \textbf{then}\  p \ \textbf{else}\ \mathsf{lca}(hp, ts) \\
				    h = ts\backslash hp \quad
					\Gamma, h \vdash (enS, \alpha_6) \xrightarrow{e}_{\mathit{enC}} (\alpha_7, b'') 
				\end{array}
			}
			{\Gamma, is  \vdash (p, (v_1, I_1)) \xrightarrow{e}_{\mathit{runS}} (\alpha_7, b'')} 
			&\\	
						&  \prftree[r]{runS3}
			{
				\begin{array}{cc}
					v_2.ev = v_1.ev[(p, e)  \mapsto v_1.ev(p,e)+1] \quad  
				v_3 = \textbf{if}\ is \ \textbf{then}\ v_2\ \textbf{else}\ v_2.tv[p \mapsto v_2.tv(p)+1] \\
				\mathsf{state}(\Gamma, p) = (p, a_i, a_d, a_e, tl_i, tl_o, C) \\
					\Gamma, p \vdash (tl_o, (v_3, I_1)) \xrightarrow{e}_{\mathit{tl}} (\alpha_2, b, \top, a_t, ts, hp)\quad b=0\vee b=-1	 \\
					\Gamma, p \vdash (a_d, \alpha_2) \xrightarrow{e}_{\mathit{a}} (\alpha_3, \top) \quad  
					\Gamma, p \vdash (tl_i, \alpha_3) \xrightarrow{e}_{\mathit{tl}} (\alpha_4, b', \top, a_t', ts', hp')\\
					b'=0\vee b'=-1	 \quad 
					\Gamma, is  \vdash (C, \alpha_4) \xrightarrow{e}_{\mathit{runC}} (\alpha_5, b'')
				\end{array}
			}
			{\Gamma, is  \vdash (p, (v_1, I_1)) \xrightarrow{e}_{\mathit{runS}} (\alpha_5, b'')} 
			&\\	
			&  \prftree[r]{runO}
			{
				\begin{array}{cc}
				\mathsf{comp}(\Gamma, p) = \mathsf{Or}(\textit{tl}, b, \textit{sf}) \quad 
			 	I_1(p) = (b', p_a, p_h) \quad
				\Gamma, is  \vdash (p_a, (v_1, I_1)) \xrightarrow{e}_{\mathit{runS}} (\alpha_2, b)	 
				\end{array}
			}
			{\Gamma, is  \vdash (p, (v_1, I_1)) \xrightarrow{e}_{\mathit{runC}} (\alpha_2, b)} 
			&\\	
			&  \prftree[r]{runSL}
			{
				\begin{array}{cc}
					\Gamma, is \vdash (s, \alpha_1) \xrightarrow{e}_{\mathit{runS}} (\alpha_2, \top) \\
					\Gamma, is\vdash (sl, \alpha_2) \xrightarrow{e}_{\mathit{runSL}} \gamma 
				\end{array}
			}
			{\Gamma, is \vdash (s\#sl, \alpha_1) \xrightarrow{e}_{\mathit{runSL}} \gamma} 
			 \quad  \prftree[r]{runA}
			{
				\begin{array}{cc}
				    \mathsf{comp}(\Gamma, p) = \mathsf{And}(\mathit{sl}, \mathit{f}) \\
					\Gamma, is \vdash (sl, \alpha_1) \xrightarrow{e}_{\mathit{runSL}} \gamma 
				\end{array}
			}
			{\Gamma, is \vdash (p, \alpha_1) \xrightarrow{e}_{\mathit{runC}} \gamma} 
			&\\							 	
		\end{eqnarray*}
	}
\caption{Semantics for state execution}
\end{figure}

Execution of a state consists of the following steps: the event and time valuations of the state are updated, taking note of the parameter $\mathit{is}$. Then the outer transitions are tried in priority order. If no outer transition succeeds, the during action of the state executes, and then the inner transitions are tried in priority order. If no inner transition succeeds, then the active substates of the state are executed. (Rule runS) defines the first case: (1) occurrences of event $e$ at state $p$ increases by 1, and the execution time of $p$ increases by 1 if it is not in the context of event handling; (2) the outer transition list $tl_o$ executes successfully, reaching the target state $ts$, with $hp$ being the lowest common ancestor during the whole transition path; (3) determine the path $\mathit{exS}$ of the state composition to exit, which is the parent of source state $p$ if the transition is from $p$ to itself, otherwise the lowest common ancestor of $p$ and $hp$, then exit the corresponding composition $\mathit{exS}$,  followed by the execution of the transition actions $a_t$; (4) determine the path of the target composition $\mathit{enS}$ to enter, and the path $h$ from the composition to the target state, and finally enter $\mathit{enS}$. The state execution completes. For the second case (rule runS2), the outer transitions failed and some inner transition succeed, so we exit and enter the compositions according to the source and target of transition (similar to the first case). For the third case (rule runS3), both the outer and inner transitions fail (denoted by the values of $b, b'$), then the composition inside the state executes.

(Rule runO) defines the  execution for $\mathsf{Or}$-composition. It first extracts the active substate of the composition via the context $I_1$ and then executes. The execution for an $\mathsf{And}$-composition executes the parallel states in the priority order and can be defined directly. The execution for an $\mathsf{And}$-composition executes the parallel states in the priority order (Rules exeA, exeSL).

\begin{example}
Revisit the example in Fig.~\ref{example:transitionpath}. When $A$ executes, it has no enabled outer or inner transitions, so its Or composition executes. $A1$ executes, then transition 1 executes first. Suppose  it successfully  reaches $A2$, then the complete transition path is found and the $hp$ for the transition path is the lca of $A1$, $A2$ and the junction, which is $A$. $exS$ and $enS$ will be $A$. According to rule (runS), the composition of $A$ exits and enters, i.e. $A1$ exits and $A2$ enters. 
But, if the junction moves outside $A$, the $hp$    becomes  the parent of $A$. $exS$ and $enS$ will be the parent of $A$. Thus, $A1, A$ exit and $A, A2$ enter in sequence.
\end{example}

\subsubsection{Semantics of a Stateflow chart}

Execution of a Stateflow chart is equivalent to execution of its top-most state composition. Given a sequence of input events $[e_1,\cdots, e_n]$, $e_i$ the trigger event at $i$-th round,  the execution of a Stateflow chart for $n$ rounds is  represented by $\Gamma \vdash \alpha_1 \xrightarrow{[e_1,\cdots, e_n]}_{\mathit{Ch}} \alpha_2$. The rule for zero round is $\Gamma \vdash \alpha \xrightarrow{[\ ]}_{\mathit{Ch}} \alpha$. Otherwise, the rule for $n>0$ rounds is as follows.

{\small 
\[
	 \prftree[r]{}
	 {
	 	\begin{array}{cc}
	 		\Gamma,  0 \vdash (root, \alpha_1) \xrightarrow{e_1}_{\mathit{runC}} (\alpha_2, \top) \quad 
	 		\Gamma \vdash \alpha_2 \xrightarrow{[e_2,\cdots, e_n]}_{\mathit{Ch}} \alpha_3	 
	 	\end{array}
	 }
	 {\Gamma \vdash  \alpha_1 \xrightarrow{[e_1,\cdots, e_n]}_{\mathit{Ch}} \alpha_3} 
 \]
}

\subsection{Determinism of the Semantics}

We prove that the above operational semantics is deterministic, as expected. The theorem in Isabelle/HOL stating determinism of semantics is given as follows. Here predicate \isa{state\_exec} corresponds to the semantic relation $\xrightarrow{e}_{\mathit{runS}}$ defined above. The theorem states that given static environment $\Gamma$, path $p$, event $e$, event handling flag $\mathit{is}$, and starting dynamic environment $\alpha=(v,I)$, if it is possible to reach dynamic environment $\alpha_1=(v_1,I_1)$ and early return flag $b_1$, as well as $\alpha_2=(v_2,I_2)$ and $b_2$, then $v_1=v_2\wedge I_1=I_2\wedge b_1=b_2$.

\begin{isabelle}
\isacommand{theorem} deterministic\_state: \\
\isasymforall st1 b1 st2 b2. state\_exec senv p e is v I v1 I1 b1 \isasymlongrightarrow \\
\isaindent{---------------------}state\_exec senv p e is v I v2 I2 b2 \isasymlongrightarrow\ v1 = v2 \isasymand\ I1 = I2 \isasymand\ b1 = b2
\end{isabelle}
The proof of this theorem mostly consists of analyzing the different cases in the operational semantics, such as actions, outer and inner transitions, state entry and exit, and so on. The full proof is over 3000 lines long.

Due to the existence of junction loops and event broadcasts, termination is not guaranteed for execution of Stateflow charts. This is also one motivation for defining our semantics as relations rather than functions.

\oomit{
\subsubsection{Mechanization}
We formalize the Stateflow language and the semantics defined above in Isabelle/HOL. Especially all the semantics are defined as inductive relations that call each other. Due to event broadcasts, the semantics of the most basic actions depends on the semantics of state and composition execution, which in turn calls the semantics of actions, transition lists, exit, and entry. 
}

\section{Automatic Execution of Stateflow Charts}
\label{sec:engine}

In this section, we present a tool for automatically executing Stateflow charts in Isabelle/HOL. This allows us to validate our semantics by testing on a large number of Stateflow charts. The automatic execution tool consists of two parts: a tactic executing the semantics in Isabelle, and a translation tool from Stateflow charts to their Isabelle representations.

\paragraph{Executable semantics in Isabelle/ML}

Automatic execution of Stateflow semantics is implemented as a tactic by writing ML code in Isabelle. Given the Stateflow chart, initial values, and a sequence of input events, it constructs an Isabelle theorem stating the result of execution according to the operational semantics. The tactic consists of functions for constructing each of the arrows in the semantics. For each arrow, the following steps are taken: first necessary inputs are collected, from which it is decided which rule should be used. Then, all premises of the rule are constructed recursively, and the rule is applied to obtain the result. 

We implemented ML functions for automatic execution of all semantic rules. This produces a final theorem corresponding to execution of a chart $\xrightarrow{el}_{\mathit{Ch}}$: 
\begin{isabelle}
\isacommand{schematic\_goal} Ch senv el denv1 ?denv 
\end{isabelle}
Here $\mathit{senv}$ and $\mathit{el}$ are the static environment and event list. $\mathit{denv1}$ is the initial value of dynamic environment, and $?denv$ represents the dynamic environment after execution, which will be constructed automatically by the tactic. For a concrete model, this goal would be solved by first expanding the definitions in the statement, followed by the tactic \textit{stateflow\_execution}.

Several optimizations in the ML code are needed to reduce its running time. First, there are several places where the same theorem need to be used multiple times. We make sure to save and reuse such theorems. Second, many steps in the derivation require simplification. Rather than using the general simplifier in Isabelle, which may be slow on large inputs, we design simplification methods that are specialized to our needs, e.g. focusing on simplification of functions and arithmetic only.



\paragraph{Translator from Stateflow to Isabelle}


In previous work, we implemented a translator from Simulink/Stateflow to representations in Python. Based on this work, our translator reads a Simulink/Stateflow model in XML format. 
Then, after calling the translator from Simulink/Stateflow to Python, it traverses the resulting Python objects of the Stateflow chart and constructs the chart according to the syntax defined in Isabelle. 

The overall architecture is shown below. The input consists of an XML file containing the Stateflow chart, and a JSON file containing execution periods, input trigger events, and expected print outputs during execution. After translation to Isabelle/HOL, the semantics is executed automatically and a theorem is produced, from which it is checked whether the output sequence is as expected. 

\vspace{2mm}
\includegraphics[width=.95\textwidth]{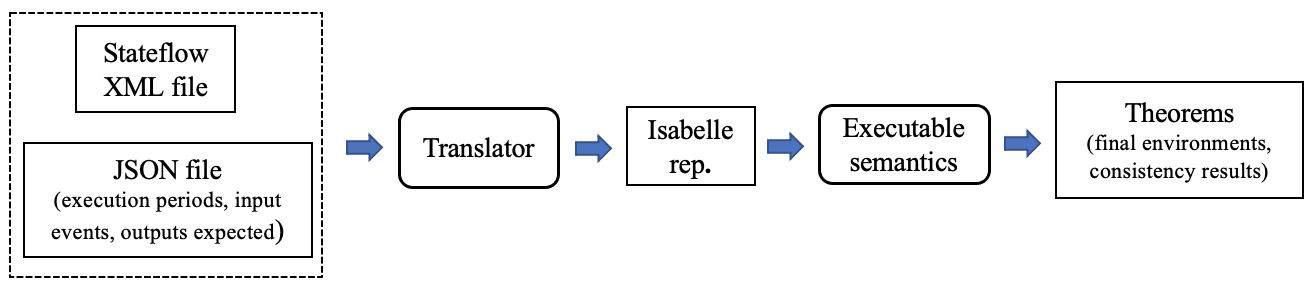}

\section{Experimental Results}
\label{sec:example}

We now discuss experiments conducted to validate our semantics, by comparing the execution results in Isabelle/HOL with simulation results in Simulink for a range of examples. We use examples from \cite[Appendix B]{Stateflow}, as well as the benchmark examples in~\cite{ExamplesBenchmark}, and some new examples designed specifically for clarifying the semantics. Over a hundred examples are tested in total. They cover all the features introduced in Section~\ref{sec:introduction}, and their execution based on our semantics is consistent with simulation. 

In addition, we test the stop-watch example from~\cite{LAP1Hamon,LAP2Hamon} and the washing machine example in Fig.~\ref{WashingMachine_Example}. For the washing machine, in order to compare the orders of execution, we insert output messages in key entry actions and transitions. Given input events \isa{START}, \isa{SWITCH}, $\varepsilon$ (10 times), \isa{SWITCH}, \isa{SWITCH}, $\varepsilon$ (33 times), where $\varepsilon$ corresponds to the cycles with no input event, the resulting theorem shows the output ``Init $\rightarrow$ 
Add Water$\rightarrow$ Washing $\rightarrow$ Pending $\rightarrow$ Washing $\rightarrow$ Add Water $\rightarrow$ Washing $\rightarrow$ Add Water $\rightarrow$ Washing $\rightarrow$ Washing Completed''. This is the expected washing cycle interrupted by one switch to \sfsyn{Pending} state.

Apart from correctness, we also test the efficiency of execution within Isabelle. The test environment is a Macbook Pro 2018, with a 2.3 GHz Intel Core i5 processor and 8GB memory. Most of the examples take 0.5s--5s for one simulation step (the washing machine example takes 3.5s for one simulation step), and a few of them with both $\mathsf{And}$-compositions and local event broadcasts take over 10s for one step. \oomit{The overall conclusion is: (1) The efficiency is positively correlated with model complexity; (2) And compositions take more time than Or compositions as we need to traverse all states in the former but only one state in the latter; (3) Occurrence of local event broadcasts will significantly increase the execution time because the chart will execute at least one extra time.}As expected, the efficiency of execution in Isabelle is  lower than Matlab/Simulink, since formal theorems must be constructed explicitly for each step taken.

\section{Conclusion}
\label{sec:conclusion}

In this paper, we defined a formal semantics of a large subset of Stateflow that covers many of its complex features, and formalized the semantics in Isabelle/HOL. Furthermore, we implemented a tool for automatic execution of the semantics starting from the Stateflow models. We validated our semantics on a number of Stateflow examples that contain various features we consider. The mechanization of the semantics and the consistency of execution results with Simulink provide strong justification for the correctness of the semantics.

The formal semantics can be used as a foundation for proving correctness of model transformations from Stateflow to other formal models. Hence, for future work, we will consider integrating this work into our model-based design framework on modelling, verification and code generation of embedded systems, from Simulink/Stateflow and AADL combined graphical models to HCSP formal models, and to implementations in SystemC or other low-level programming languages. This semantics is intended to be used in a machine-checked proof for correctness of translation between Stateflow and HCSP programs, which when combined with techniques for verifying HCSP programs, allows to formally verify correctness and safety properties of Stateflow models. 

\paragraph{Acknowledgements.} 
This work is supported in part by the NSFC under grants No. 61972385, 61732001 and 62032024.

\bibliographystyle{splncs04}
\bibliography{ref}

\begin{appendix}

\section{Executable Semantics in Isabelle}

In this section, we give some details about implementation of automatic execution of Stateflow semantics by writing tactics in Isabelle/ML.

We use the case of state execution to illustrate this process. When (rule runS) is applied, it results in a theorem of the form $\Gamma, is \vdash (p, \alpha_1) \xrightarrow{e}_{\mathit{runS}} (\alpha_2, b)$ for some dynamic environment $\alpha_2$. In the implementation, the corresponding ML function is $\mathsf{evaluate\_outer\_trans\_s}$. The function first instantiates the initial environments in the semantic rule, and then calculates and resolves the premises in sequence.  

\begin{isaenv}
    $\mathsf{evaluate\_outer\_trans\_s}$ $\mathit{senv}$ $p$ $e$ $is$ $\mathit{denv1}$ = 
    let
        $\mathit{th}1$ = $@\{$thm outer_trans_semantics$\}$
        $\mathit{inst}$ = $\cdots$ // $\mbox{extract the mappings}$ 
        $\mathit{th}2$ = $\mathit{th1}$ |> $\mathsf{Drule.instantiate\_normalize}\ \mathit{inst}$
        $\mathit{denv2}$ = $\cdots$ // update valuations
        $\mathit{th}3$ = ($\mathsf{evaluate\_tl}$ $senv\ tl_o\ e\ p\ \mathit{denv2})\ \mathsf{RS}\ \mathit{th}2$
        $\mathit{exit\_p}$ = ... // get the composition path to$\mbox{ exit}$ from
        $\mathit{th}4$ = ($\mathsf{evaluate\_exit\_C}\ senv\ $ $exit\_p\ e\ \mathit{denv3})\ \mathsf{RS}\ th3$
        $\mathit{th}5$ = ($\mathsf{evaluate\_actionlist}\ senv\ a_t\ p\ \mathit{denv4})\ \mathsf{RS}\ th4$
        $h$, $\mathit{entry_p}$ = ... // get the target paths
    in
        ($\mathsf{evaluate\_entry\_C}\ \mathit{senv}\ $ $\mathit{entry_p}\ e\ h\ \mathit{denv5})\ \mathsf{RS}\ th5$
    end
\end{isaenv}


The parameters of the function include the static environment $\mathit{senv}$, path $p$ to be executed (from which the state $s = (p, a_i, a_d, a_e, tl_i, tl_o, C)$ is obtained), triggering event $e$, event handling flag $\mathit{is}$, and initial dynamic environment $\mathit{denv1}$. 
The implementation can be understood as follows. First, obtain theorem $\mathit{th}1$ corresponding to the semantics of outer transition, extract the instantiation mappings $\mathit{inst}$, and then instantiate the schematic variables in $\mathit{th}1$ with $\mathit{inst}$ using built-in function $\mathsf{Drule.instantiate\_normalize}$, which returns a concrete theorem with a sequence of premises. Next, calculate the new environment $\mathit{denv2}$ by increasing event occurrences and time of the state, obtain the outer transition theorem by calling $\mathsf{evaluate\_tl}$ with corresponding arguments, and then resolve $\mathit{th}2$ by eliminating the premise corresponding to $\mathsf{evaluate\_tl}$ to get a new theorem $\mathit{th}3$.
Following the steps in semantics, next calculate the paths and compositions to exit, and then exit the corresponding composition in $\mathit{th}4$, execute the transition actions in $\mathit{th}5$, and calculate the paths and compositions to enter, and finally enter the target composition. The state execution completes and returns the final theorem.

Similarly, we can define the functions for executing other semantic rules corresponding to inner transitions and composition of state execution, respectively. Next we give the structure for the definition for executing a state:
\begin{isaenv}
    $\mathsf{evaluate\_s}$ $\mathit{senv}$ $p$ $e$ $is$ $\mathit{denv1}$ = 
    let
        $\mathit{denv2}$ = $\mathsf{incr}$($denv1$, $p$, $e$, $is$) // get new environment
        $\mathit{outer\_trans\_th}$ = $\mathsf{evaluate\_tl}$ $\mathit{senv}$ $tl_o$ $e$ $p$ $\mathit{denv2}$
        $vt1$, $\mathit{denv3}$ = ... // get transition status and new environment
    in
        if $vt1$ = 1 then
            $\mathsf{evaluate\_outer\_trans\_s}$ $senv$ $p$ $e$ $is$ $\mathit{denv1}$
        else 
            $\mathit{inner\_trans\_th}$ = $\mathsf{evaluate\_tl}$ $senv$ $tl_i$ $e$ $p$ $\mathit{denv3}$
            $vt2$ = ... // $\mbox{get transition status}$
            if $vt2$ = 1 then
                $\mathsf{evaluate\_inner\_trans\_s}$ $senv$ $p$ $e$ $is$ $\mathit{denv1}$
            else
                $\mathsf{evaluate\_comp\_s}$ $senv$ $p$ $e$ $is$ $\mathit{denv1}$
    end
\end{isaenv}

As shown above, we first evaluate and execute the outer transition list $tl_o$ by calling $\mathsf{evaluate\_tl}$, from which the transition status $vt1$ is obtained. If $vt1$ is equal to 1 (indicating that some outer transition succeeds to execute), then $\mathsf{evaluate\_outer\_trans\_s}$ is called from initial dynamic environment $\mathit{denv1}$; otherwise, the inner transition list $ti$ is executed and the transition status $vt2$ is obtained. If $vt2$ is 1, then $\mathsf{evaluate\_inner\_trans\_s}$ is called from $\mathit{denv1}$, otherwise, $\mathsf{evaluate\_comp\_s}$ is called for executing the composition, which corresponds to rule (runS2) and (runS3) presented in Appendix A.5.

	
	
Following the above processes, we implement all the semantic rules in ML, especially $\mathsf{evaluate\_C}$ is defined for the automatic execution of a Stateflow chart. We define a tactic which calls the main function $\mathsf{evaluate\_C}$ for automatically executing a chart, to build a theorem corresponding to the above goal. 
 
\begin{isaenv}
    $\mathsf{stateflow\_execution\_tac}$  $state$ = 
    let
        $\mathit{subgoals}$ = $state$ |> $\mathsf{Thm.cprop\_of}$ |> $\mathsf{Drule.strip\_imp\_prems}$
    in
        if $\mathsf{null}$ $\mathit{subgoals}$ then $\mathsf{Seq.empty}$ else
            ... // get all arguments from $state$ 
            $th$ = $\mathsf{evaluate\_C}$ $senv$ $root$ $e$ $is$ $denv$
            $\mathsf{Seq.single}$ ($th$ $\mathsf{RS}$ $state$)
    end
\end{isaenv}
where $\mathsf{evaluate\_C}$ is called to build the corresponding theorem $th$ for a non-empty goal. 
 
\end{appendix}

\end{document}